\documentclass[preprint,10pt]{elsarticle}

\usepackage{graphicx}
\usepackage{amssymb}
\newcounter{bla}

\journal{Computer Physics Communications}

\begin{document}

\begin{frontmatter}



\title{GR@PPA 2.9: radiation matching for simulating photon production 
processes in hadron collisions}


\author{Shigeru Odaka\corref{odaka}}
\author{Yoshimasa Kurihara}

\cortext[odaka] {Corresponding author.\\
\textit{E-mail address:} shigeru.odaka@kek.jp}

\address{High Energy Accelerator Research Organization (KEK)\\
1-1 Oho, Tsukuba, Ibaraki 305-0801, Japan}

%--------1---------2---------3---------4---------5---------6---------7---------8
\begin{abstract}
We release an event generator package, GR@PPA 2.9, for simulating the direct 
(single) photon and diphoton (double photon) production in hadron collisions.
The included programs were used in our previous studies, in which we have 
explicitly shown large contributions from parton-associated processes.
The programs consistently combine simulations based on matrix elements
with parton-shower simulations that reproduce the multiple parton radiation 
and quark fragmentation to photons.
The matrix elements include associated parton production processes 
up to two partons.
We provide instructions for the installation and execution of the programs 
in this article.
The practical performance is also presented.
\end{abstract}


\end{frontmatter}

% \linenumbers

% Computer program descriptions should contain the following
% PROGRAM SUMMARY.

\noindent
{\bf PROGRAM SUMMARY}
  %Delete as appropriate.

\begin{small}

\noindent
{\em Manuscript Title:} GR@PPA 2.9: radiation matching for simulating 
photon production processes in hadron collisions \\
{\em Authors:} S. Odaka, Y. Kurihara\\
{\em Program Title:} GR@PPA 2.9\\
{\em Journal Reference:}                                      \\
  %Leave blank, supplied by Elsevier.
{\em Catalogue identifier:}                                   \\
  %Leave blank, supplied by Elsevier.
{\em Licensing provisions:} none\\
  %enter "none" if CPC non-profit use license is sufficient.
{\em Programming language:} Fortran; 
with some included libraries coded in C and C++\\
{\em Computer:} all\\
  %Computer(s) for which program has been designed.
{\em Operating system:} any UNIX-like system\\
  %Operating system(s) for which program has been designed.
{\em RAM:} 4 Maga bytes\\
  %RAM in bytes required to execute program with typical data.
% {\em Number of processors used:} \\
  %If more than one processor.
{\em Supplementary material:}                                 \\
  % Fill in if necessary, otherwise leave out.
  % Please give some freely chosen keywords that we can use in a
  % cumulative keyword index.
{\em Classification:} 11.2\\
  %Classify using CPC Program Library Subject Index, see (
  % http://cpc.cs.qub.ac.uk/subjectIndex/SUBJECT_index.html)
  %e.g. 4.4 Feynman diagrams, 5 Computer Algebra.
{\em External routines/libraries:} bash and Perl for the setup, 
and CERNLIB, ROOT, LHAPDF 5, PYTHIA 6.4 according to the choice of users\\
  % Fill in if necessary, otherwise leave out.
%{\em Subprograms used:}                                       \\
  %Fill in if necessary, otherwise leave out.
% {\em Catalogue identifier of previous version:}*              \\
  %Only required for a New Version summary, otherwise leave out.
% {\em Journal reference of previous version:}*                  \\
  %Only required for a New Version summary, otherwise leave out.
{\em Does the new version supersede the previous version?:}
No, this version supports processes not included 
in previous versions.\\
{\em Nature of problem:}
It is necessary to include jet-associated processes up to two jets 
for realistic simulations of photon production processes in high-energy 
hadron collisions.
Photons also need to be considered as radiation.
An appropriate matching method has to be introduced for combining 
those processes having different jet and photon multiplicities and 
parton shower simulations.
\\
{\em Solution method:} 
The limited leading-log (LLL) subtraction method has been extended to 
multi-jet processes and those including photon radiation.
The final-state parton shower simulation has been improved 
to support QED photon radiation.
\\
{\em Reasons for the new version:}
New processes have been supported.\\
{\em Summary of revisions:}
This program package provides event generators for realistic simulations 
of single and double photon production processes in high-energy 
hadron collisions.
Users can consistently combine different jet- and photon-multiplicity 
processes up to two jets and two photons.
\\
{\em Restrictions:} 
\\
{\em Unusual features:} 
\\
%{\em Additional comments:}\\
{\em Running time:} 
The CPU time consumption is strongly dependent on the process.
An example is presented in the text for a sample program.
\\
%--------1---------2---------3---------4---------5---------6---------7---------8

\end{small}

\section{Introduction}
\label{sec:intro}

High-energy isolated photon production is a clean signal 
in busy final states of hadron collisions 
and is considered to be suitable for probing the internal structure 
of the colliding hadrons~\cite{Halzen:1979pb,Owens:1986mp}, 
as such photons are expected to be produced via well-known Quantum 
Electrodynamic (QED) interactions of constituant partons 
(gluons and quarks), 
which is referred to as the direct photon production. 

Despite such a prospect, this process has been a mystery for a long time.
Theories, even the next-to-leading order (NLO) calculations 
in Quantum Chromodynamics (QCD), could not reproduce measurement results, 
and intensive studies and debates were made to solve the problem 
(see, for example,~\cite{Kumar:2003kg,Aurenche:2006vj} 
and the references therein).
As higher energy collision data became available~\cite{Abazov:2005wc,
Aaltonen:2009ty,Aad:2011tw,Chatrchyan:2011ue}, it has been revealed 
that the photon production in high transverse momentum ($p_{T}$) regions 
($\gtrsim 50$ GeV) can be reasonably described 
by the NLO calculation~\cite{Catani:2002ny}, 
although discussions are yet to be conclusive for lower $p_{T}$ production.

From the discussions, it has been recognized that appropriate treatments 
of divergent collinear contributions in both QCD and QED, 
the multiple radiation in QCD, and the fragmentation in QED, are important.
Possible problems in the definition of the isolation condition have also 
been addressed~\cite{Frixione:1998jh}.
In addition, although it has not been emphasized explicitly, 
the NLO correction that includes the effect of associated two-parton 
production was found to be unreasonably large in the sense of perturbation.
The correction amounts to more than 100\% under typical measurement 
conditions~\cite{Catani:2002ny}\footnote{Recently, the NNLO correction 
to the direct photon production was found to be moderate~\cite{Campbell:2016lzl}.}.

Along with the direct (single) photon production, 
diphoton (double photon) production is also an important process, 
especially in searches for new phenomena.
Indeed, the diphoton production at the Large Hadron Collider (LHC) 
played an important role in the discovery 
of the Higgs boson~\cite{Aad:2012tfa,Chatrchyan:2012xdj}.
In the diphoton mode, 
the Higgs boson was observed as a small resonance on a large 
background from other interactions.
Hence, the understanding of the non-resonant background is crucial 
for detailed studies on the discovered Higgs boson properties.

The diphoton production also shows a dramatic and mysterious behavior.
First of all, the NLO correction is markedly large; it is more 
than 200\% under typical measurement conditions~\cite{Binoth:1999qq}.
This is predominantly caused by a very large contribution of quark-gluon 
($qg$) interactions that newly participate in this order.
Furthermore,  
it was recently found that the next-to-next-to-leading order (NNLO) 
correction that includes the effects of associated two-parton production 
is also large, 
as in the direct photon production~\cite{Catani:2011qz,Campbell:2016yrh}.
Hence, this order of correction is necessary to be included 
in order to realistically reproduce the non-resonant 
diphoton production~\cite{Aad:2012tba,Chatrchyan:2014fsa}, although we still 
do not have any reasonable explanation on this large correction.
Here, we note that the large NNLO correction is not predominantly caused 
by the emergence of new processes, 
{\it e.g.}, $gg \rightarrow \gamma \gamma q\bar{q}$, 
but mainly originates from ordinary gluon-radiation corrections to 
one-parton production processes~\cite{Catani:2011qz}.

In this article, 
we describe a program package, GR@PPA 2.9, for simulating the direct photon 
and diphoton production in proton-proton ($pp$) 
and proton-antiproton ($p\bar{p}$) collisions.
The included event generator programs were used for the studies reported 
in our previous articles~\cite{Odaka:2012ry,Odaka:2015uqa,Odaka:2016tef}, 
in which we confirmed large contributions from parton-associated processes. 
The predominance of gluon-radiation processes among two-parton processes 
was also found in the direct-photon production~\cite{Odaka:2016tef}.

In GR@PPA 2.9, 
the event generation is based on leading order (LO) matrix elements (MEs) 
for the processes including associated parton production up to two partons.
The programs for calculating the MEs were produced by using the GRACE 
system~\cite{Ishikawa:1993qr,Yuasa:1999rg}.
The predominant collinear corrections of QCD, 
which lead to the multiple radiation of partons, 
are simulated by applying parton shower (PS) simulations to the generated 
events~\cite{Odaka:2007gu,Odaka:2009qf,Odaka:2011hc,Odaka:2012ry}.
The simulation can proceed further to the hadron level with the help of 
external general-purpose event generators, 
such as PYTHIA~\cite{Sjostrand:2006za}.

The QCD corrections are involved in both ME-based event generation and PS.
The matching between the two simulations is achieved 
using the limited leading-log (LLL) subtraction 
method~\cite{Kurihara:2002ne,Odaka:2007gu,Odaka:2012ry}, 
in which the divergent leading-logarithmic (LL) components are numerically 
subtracted from squared MEs of radiative processes to make them finite.
The LL components are the leading terms of PS.
We achieve good matching by carefully arranging the PS kinematics and 
energy scale matching, to combine the events that are generated 
according to the MEs of different parton multiplicities\footnote{ 
Other approaches to the matching in photon production processes 
can be found in~\cite{Hoeche:2009xc,DErrico:2011cgc}.}.

Along with the collinear divergence, 
we also need to consider the soft-gluon divergence to make 
the two-parton production processes finite.
We have introduced a combined subtraction method for subtracting 
these divergences simultaneously~\cite{Odaka:2014ura}.
A small correction is applied to the PS-applied lower parton-multiplicity 
events in order to compensate for this alteration in the subtraction.

In the measurements of photon production, 
we cannot completely avoid QED collinear divergences 
even if isolation requirements are imposed on the photons.
We may miss the hadronic activities induced by the quark 
that radiated the detected photon, 
if the remaining quark momentum after the radiation is small.
We apply the QED LLL subtraction to regularize such QED collinear 
divergences~\cite{Odaka:2012ry}.
The PS simulation involves QED photon radiation for restoring 
the subtracted collinear components.
In addition, very small $Q^{2}$ regions that the PS simulations do not cover 
are simulated on the basis of a fragmentation function 
(FF)~\cite{Bourhis:1997yu}.
This PS/FF simulation has the ability to enforce the radiation 
of a given number of energetic photons for efficient event 
generation~\cite{Odaka:2012ry,Odaka:2016tef}.

We can combine the events of processes involving associated partons 
in the final state up to two partons using the programs in GR@PPA 2.9.
Therefore, the approximation order is the same as NLO 
in the direct photon production and NNLO in the diphoton production.
However, finite components in loop corrections and next-to-leading 
logarithmic (NLL) components are missing.
Accordingly, the overall normalization is not theoretically guaranteed.
In contrast, transverse activities induced by the QCD multiple radiation, 
which are ignored in perturbative calculations, 
are reproduced by the PS in our simulation.
We can finally obtain exclusive hadron-level event information.
These features allow us to perform detailed studies on the acceptance, 
isolation conditions, and associated hadron-jet production 
which cannot be carried out using perturbative calculations.

The simulation method and the resultant physics performance have already been 
presented in previous articles~\cite{Odaka:2012ry,Odaka:2015uqa,Odaka:2016tef}.
We present only the execution instruction and practical performance of 
the programs in this article.
The remainder of this article is organized as follows.
The instructions for building the libraries and sample programs are 
provided in Section~\ref{sec:instruction}.
Section~\ref{sec:control} describes the control parameters 
that are newly introduced in this version.
The practical performance of a sample program is presented 
in Section~\ref{sec:perf}, 
and the descriptions are summarized in Section~\ref{sec:summary}.

\section{Instruction}
\label{sec:instruction}

\subsection{Distribution package}

The program package of GR@PPA 2.9 is distributed as a gzipped tar file 
named {\tt GR@PPA-2.9.tgz}, 
which can be obtained from the GR@PPA Web page\footnote{\tt
http://atlas.kek.jp/physics/nlo-wg/grappa.html\#GRAPPA2.9}.
The compressed file can be expanded, for instance, by typing
\renewcommand{\baselinestretch}{0.78}
\begin{verbatim}
    tar zxf GR@PPA-2.9.tgz
\end{verbatim}
on UNIX systems.
When the file is expanded, 
users have a directory named {\tt GR@PPA-2.9} containing 
the following files and directories:

\begin{tabbing}
(miscellaneous files)\\
{\tt README} $\quad \quad \quad$ \= : \= readme file describing how to use 
this package,\\
{\tt VERSION-2.9} \> : \> file to show the version number,\\
\\
(files for setup)\\
{\tt config.input} \> : \> file to specify the configuration 
for the setup,\\
{\tt Install} \> : \> shell script for the installation,\\
{\tt config} \> : \> shell script to configure the setup,\\
{\tt config.perl} \> : \> Perl script called by {\tt config},\\
{\tt proc.list} \> : \> process list, which is referred to 
in {\tt config.perl},\\
\\
(GR@PPA framework)\\
{\tt grckinem} \> : \> source files of the framework,\\
{\tt basesv5.1} \> : \> BASES 5.1 source package with some customization,\\
{\tt chanel} \> : \> CHANEL source package to define the interaction model,\\
{\tt inc}  \> : \> directory containing common include files,\\
{\tt example} \> : \> directory to be used for the setup of sample programs,\\
{\tt diagrams} \> : \> directory containing PS files illustrating typical 
Feynman diagrams,\\
{\tt lib} \> : \> directory to store object libraries; initially empty,\\
{\tt ffphoto} \> : \> QED fragmentation function,\\
\\
(process directories)\\
{\tt ajets} \> : \> single-photon production processes,\\
{\tt aajets} \> : \> double-photon production processes,\\
{\tt qcd} \> : \> QCD-jet production processes,\\
{\tt handcode} \> : \> handcoded ME programs.
\end{tabbing}

The setup files and framework directories are almost identical to those 
in the GR@PPA 2.8 series~\cite{Odaka:2011hc}.
The {\tt ffphoto} directory has been added to simulate the small $Q^{2}$
photon radiation according to an FF~\cite{Bourhis:1997yu}.
All modifications to the framework routines are now backported to 
recent versions of the 2.8 series, {\it e.g.}, GR@PPA 2.8.6.2.

The process directories are entirely different from those in the 2.8 series.
The directories {\tt ajets} and {\tt aajets} contain ME calculation programs 
for single and double photon production processes, respectively.
The {\tt qcd} directory contains the programs for QCD 2-jet production, 
in which photons can be produced if energetic photons are radiated 
in the PS simulation.
The {\tt handcode} directory contains handcoded ME programs.
Programs for the loop-mediated $gg \rightarrow \gamma\gamma$ process 
are provided in the present version.

\subsection{Installation}

The installation procedure is also identical to that for 
the GR@PPA 2.8 series.
Move into the expanded distribution package, 
and appropriately edit the file {\tt config.input} 
to specify the parton distribution function (PDF) library, 
compilation tools, and the location of external libraries.

For the PDF library, users have to select one of the following options:  
{\tt CTEQL1}, {\tt LHAPDF}, or {\tt PDFLIB},
although the option {\tt PDFLIB} is now obsolete 
because programs built with this option refer to the old PDFLIB 
library in CERNLIB.
The programs built with {\tt CTEQL1} use the CTEQL1 PDF~\cite{Pumplin:2002vw} 
contained in this package.
This choice is convenient for a quick check of the program.
The option {\tt LHAPDF} should be selected for intensive studies.
The programs built with this option refer to an LHAPDF 
library\footnote{We have been using the LHAPDF version 5 for our studies. 
The compatibility with the version 6 has not been tested.}~\cite{Whalley:2005nh}.
If users select this option, they have to prepare an external LHAPDF 
library and specify its location in {\tt config.input}.

External libraries are necessary for building sample programs.
Although they can be modified later, 
it is recommended to specify them in {\tt config.input} 
to ensure the consistency.
Here, unnecessary libraries are allowed to be left unspecified.

The libraries are compiled and installed in the {\tt lib} directory 
by executing 
\renewcommand{\baselinestretch}{0.78}
\begin{verbatim}
    ./Install
\end{verbatim}
at the top of the expanded distribution package.
It may be better to separately execute the two commands, 
{\tt ./config} and {\tt make}, when certain problems are encountered.
More details about the installation can be found 
in the GR@PPA-2.8 manual~\cite{Odaka:2011hc}.

\subsection{Sample programs}

\subsubsection{Built-in samples}

Sample programs are provided under the {\tt ajets/example} 
and {\tt aajets/example} directories for a quick check of the installation.
Programs using both HBOOK and RBOOK are provided 
as in the GR@PPA-2.8 distributions.
Users can build and execute the program by invoking the commands 
\renewcommand{\baselinestretch}{0.78}
\begin{verbatim}
    ./Config
    make
    ./run
\end{verbatim}
in one of the sample program directories.
The sample programs produce an event file using the LhaExt 
utility\footnote{{\tt http://atlas.kek.jp/physics/nlo-wg/grappa.html\#LhaExt}}.
The version 2.0 of LhaExt is implemented in GR@PPA 2.9 and 
recent versions of GR@PPA 2.8, 
in which the XML-based LHEF format of event data~\cite{Alwall:2006yp} 
is supported.

The event data can be further processed down to the hadron level 
with the help of PYTHIA 6.4~\cite{Sjostrand:2006za} by executing 
the program provided under the {\tt pythia} directory in each sample.
After the hadronization, 
the final angular and momentum selection is applied and 
the isolation requirement is imposed to the photons in these programs.
The parameters in PYTHIA 6.4 are left unchanged from the default in the sample 
programs, except for {\tt PARP(67) = 1.0} and {\tt PARP(71) = 1.0}, 
which are explicitly specified in the source program {\tt Pythia.f}.
This setting should be applied when the default "old" PS is used in PYTHIA.
Users can build and execute the sample programs by invoking the above three 
commands again in the {\tt pythia} directory.

Contrary to heavy particle production, 
in which no-cut event generation is possible, 
strict angular and momentum constraints are necessary to be imposed 
in photon production processes.
The constraints need to be applied at several stages of the simulation 
in order to optimize the event generation efficiency.
In the sample programs, 
constraints are imposed at the generation of parton-level hard-interaction 
events, after applying the PS simulations, 
and finally after completing the hadronization.
Since only the final selection at the hadron level is physically meaningful, 
we need to confirm that the selected sample is not biased 
by the constraints of earlier stages.

In order to make the above confirmation possible, the sample programs 
are arranged in such a way that the initial hard-interaction 
information can be added to the event information that is passed to 
the PYTHIA application program.
The added information is separated from the ordinary event information 
in the {\tt UPEVNT} subroutine attached to {\tt Pythia.f}, 
so that the information can be referred in other parts of the program.
The earlier constraints are safe in the final event sample 
if the distributions of the used quantities vanish before reaching 
the applied cuts.

\subsubsection{Matched direct photon and diphoton + two jets}

Program sets for simulating the direct photon and diphoton production 
in 7-TeV proton-proton collisions can be obtained from the GR@PPA Web page.
Associated jet production is included up to two jets in these samples.
Here, {\it jet} refers to the produced parton 
because energetic parton production is considered to be observed as 
the production of a hadron jet.
The programs were used to simulate the ATLAS measurements 
in previous articles~\cite{Odaka:2015uqa,Odaka:2016tef}.
In these external samples, the events of different photon and parton 
multiplicities are generated separately.
Such a procedure is convenient because the program setting 
and required CPU time are markedly different between the subprocesses.

As these programs are separated from the distribution package, 
{\tt Makefile} and the {\tt usr} directory have to be manually copied 
from one of the properly-built built-in sample programs described above.
The file {\tt lhapdf\_setup.sh} and the symbolic link {\tt PDFsets} 
have also to be copied when LHAPDF is selected for the PDF library.
If these files and directories are copied correctly, 
the programs can be built by simply invoking the {\tt make} command 
and can be executed by typing {\tt ./run}.
These samples contain a PYTHIA application program in each subprocess 
directory for checking the generation and selection conditions 
in the GR@PPA event generation.

Another PYTHIA application program is provided outside the event generation 
directories in these samples.
These programs have a function of mixing separately generated event data.
The mixing is carried out randomly in proportion to the integrated 
cross section of each subprocess, so that the obtained results can be 
directly compared with measurement results.
A sample {\tt Makefile} is provided because the I/O scheme is unusual 
in these programs.
The event files to be read are explicitly specified in the subroutine 
{\tt upinit} included in {\tt Pythia.f}.

The execution results of these sample programs with the use of the 
CTEQ6L1 PDF are provided under the {\tt results} directories.
The BASES result files and log files are provided for event generation 
programs, and the log files are provided for PYTHIA application programs.
These sample results will be useful for the users to confirm the execution.

\section{Control parameters}
\label{sec:control}

As can seen in the sample programs, 
users are allowed to modify the programs in four files, {\tt grappa.f}, 
{\tt upinit.F}, {\tt upevnt.f}, and {\tt grcpar.F} 
for the GR@PPA event generation.
Among them, {\tt upevnt.f} must be less attractive because it simply 
calls the steering routine of GR@PPA for generating one event.
The generation conditions are specified in {\tt upinit.F} and 
{\tt grcpar.F}.
The user analyses and data I/O are controlled in the main program in 
{\tt grappa.f}.
In the following subsections, we describe how various execution conditions 
that have been introduced after the release of GR@PPA 2.8
can be specified in these files.
Refer to the GR@PPA 2.8 manual~\cite{Odaka:2011hc} 
for the other functions of GR@PPA.

\subsection{{\tt upinit.F}}

The file {\tt upinit.F} contains only one subroutine {\tt UPINIT}, 
which is called only once at the beginning of the event generation.
Thus, the conditions that are valid through the execution are defined 
in this routine.
New or extended parameters that are supported after the release of 
GR@PPA 2.8 are summarized in Table~\ref{tab:upinit}.

\begin{table}[t]
\caption{New or extended parameters to be defined in {\tt UPINIT}. 
"D" denotes the default setting.}
\label{tab:upinit}
\begin{center}
\begin{tabular}{l l} 
\hline
Parameter & Description \\ 
\hline
{\tt labcut} & switch for activating the LabCut framework. (D = 0)\\
 & = 0 : deavtivate. \\
 & = 1 : activate. \\
\\
{\tt ishwfin} & switch for the final-state PS. (D = 0)\\
 & = 0 : no PS is applied. \\
 & = 1 : QCDPSf is applied. \\
 & = 2 : QCDPSf including QED FSR is applied. \\
 & = 3 : QCDPSf with forced QED FSR is applied. \\
\\
{\tt isgcorr} & switch for activating the soft-gluon correction. (D = 0)\\
 & = 0 : deactivate. \\
 & = 1 : activate. \\
 \hline
\end{tabular}
\end{center}
\end{table}

The LabCut framework is activated by setting {\tt labcut = 1} 
in all the sample programs.
In this framework, PS simulations are applied before passing 
the differential cross section value to the integration and event generation 
utility, BASES/SPRING~\cite{Kawabata:1985yt,Kawabata:1995th}.
The event weight that is evaluated after the application of PS can be 
reflected by the cross section integration and generation optimization.
The LabCut framework should be deactivated ({\tt labcut = 0}) 
if it is not necessary, 
because the PS application consumes additional CPU time 
for the integration.

The inclusion of QED photon radiation in the final-state PS (QED FSR) 
is controlled by the {\tt ishwfin} parameter.
The QED radiation is simulated in a democratic way, 
together with the QCD radiation, when {\tt ishwfin = 2}.
If it is specified as {\tt ishwfin = 3}, the QED radiation is enforced 
and users have to give the number of photons to be generated ({\tt nphgen}) 
and the minimum photon energy in the center-of-mass frame ({\tt ephmin})
explicitly in {\tt UPINIT}.
The {\tt ephmin} value should be equal to the minimum requirement of 
the photon transverse momentum.

In addition to the collinear divergences, 
we have to take the soft-gluon divergence into consideration in order to 
regularize the MEs of the processes including two partons in the final state.
We apply a combined subtraction in which both divergences are simultaneously 
subtracted.
Whereas, the PS simulations for restoring the subtracted components 
include the collinear components only. 
We need to apply a correction to PS-applied lower parton-multiplicity events 
in order to compensate for the alteration in the subtraction.
The activation of this correction is controlled 
by the parameter {\tt isgcorr}.

The event weights of the forced photon radiation and soft-gluon correction 
can be evaluated only after the PS simulations are applied.
Therefore, the LabCut framework has to be activated ({\tt labcut = 1}) 
when one of these functions is activated.

\subsection{{\tt grcpar.F}}

The routines in {\tt grcpar.F} are frequently called during the execution.
The subroutine {\tt GRCPAR} in {\tt grcpar.F} controls the execution of 
BASES.
The parameter {\tt NCALL} defines the accuracy of the integration.
We usually recommend that the {\tt NCALL} value should be increased 
until the integration accuracy is reduced down to 0.2\%.
However, it is difficult to achieve this requirement for large multiplicity 
processes such as diphoton + two jets.
Although better integration accuracy leads to a better event generation 
efficiency, worse integration accuracy does not deteriorate the accuracy 
of event generation, unless unreasonable jumps of the integration result 
are observed in the iteration steps 
or the event weight exceeds the estimated maximum during event generation.
Users should find a reasonable {\tt NCALL} value for minimizing the total 
CPU time, although an integration accuracy of better than 1\% is desired 
to avoid unexpected behaviors.

Another important function of the routines in {\tt grcpar.F} is to determine 
the energy scales of the events.
It is recommended to use the user-defined definition by setting 
{\tt ICOUP = 6} and {\tt IFACT = 6} in {\tt GRCPAR} 
when the radiation matching is required.
The subroutine {\tt GRCUSRSETQ} is called for defining the renormalization 
scale {\tt GRCQ} and the factorization scale {\tt GRCFAQ} with this setting.
However, this scheme is not sufficient to accomplish satisfactory matching 
because multiple definition is not allowed.
We need to evaluate the energy scales for embedded non-radiative events 
during the subtraction, simultaneously with those for the radiative event 
under consideration.
In addition, sometimes we may have to define the PS and subtraction energy 
scales independently of the renormalization and factorization scales.

The routines in the sample programs are arranged in order to fulfil 
these requirements.
The base definition is given by the subroutine {\tt setscale} according 
to the given set of four-momenta of interacting particles.
This routine is called by {\tt GRCUSRSETQ} for setting {\tt GRCQ} and 
{\tt GRCFAQ}, and it is also called by matching routines for tentatively 
evaluating the energy scales of non-radiative events.
The routine for defining the final-state PS scale ({\tt scale\_fsr}) 
and that for defining the final-state subtraction scale ({\tt scale\_fsub}) 
are also separately provided.
Although these two scales should be identical to accomplish the matching, 
independent routines are prepared because users may want unusual 
definitions for complicated studies.
The PS and subtraction scales in the initial state are always 
identical to the factorization scale in our matching method.

The selection of generated events can be done by two subroutines, 
{\tt GRCUSRCUT} and {\tt grclabcut}.
The selection in {\tt GRCUSRCUT} is carried out on the basis of 
the hard-interaction event information before applying PS simulations, 
while the selection in {\tt grclabcut} is based on the information 
after the application of PS.
The latter is called before passing the cross section value to BASES/SPRING 
when the LabCut framework is activated by {\tt labcut = 1}.
In this case, the selection condition in {\tt grclabcut} is reflected 
by the cross section integration and event-generation optimization.
In the sample programs, the event selection is completed by {\tt grclabcut} 
and no further selection is applied.
Hence, the integrated cross section of BASES can be referred to 
in the event mixing without any correction.
We recommend users to use {\tt grclabcut} for the event selection, 
when the selection is based on quantities that may be strongly affected 
by the application of PS simulations, 
such as the transverse momenta and production angles of photons 
in photon production processes.
Here, we note that {\tt grclabcut} is identical among all the subprocesses 
in the sample programs, 
while the hard-interaction generation condition is different.

The soft-gluon correction needs to be applied to lower parton-multiplicity 
processes in order to compensate for the alteration of the subtraction 
caused by the existence of soft-gluon divergences 
in multiple parton production processes.
Although the overall activation of the correction is controlled by the 
{\tt isgcorr} parameter in {\tt UPINIT}, 
users may want a process-dependent control 
when events of multiple processes are simultaneously generated.
Such a control can be achieved by using the subroutine {\tt sgcreq}.
This routine is called only when {\tt isgcorr = 1}.

\subsection{{\tt grappa.f}}

This file contains the main program for the event generation, 
in which the overall event-generation control, event analyses, and 
I/O control are carried out.
The I/O routines in GR@PPA 2.9 supports three formats of the output 
event file: ASCII, binary, and LHEF.
The selection can be done by using the {\tt ifmt} parameter.

The sample programs also have a function to add hard-interaction 
event information to the output events.
This information is useful for checking the possible bias of 
the event generation condition defined in {\tt UPINIT} and 
{\tt GRCUSRCUT} to the finally selected sample.
Users need to appropriately modify the programs that read 
the recorded events in order to utilize the added information.
This function should be deactivated if such a modification cannot be done.

\section{Practical performance}
\label{sec:perf}

The portability of the program has been tested on various Linux 
distributions.
The programs have been developed on Scientific Linux CERN (SLC) 5, 
and the scripts for building and executing the programs have been tailored 
so that all the functions should work properly on SLC 6 as well.
The used compilers are gcc 4.1 on SLC 5 and gcc 4.8 on SLC 6.
The execution of the built-in sample programs was also tested on more recent 
Linux distributions: CentOS 7 (gcc 4.8), Ubuntu 16.04 (gcc 5.4), 
and Fedora 24 (gcc 6.3).
We confirmed that we can build and execute the sample programs 
if external libraries are properly installed.
The PYTHIA and LHAPDF libraries have to be installed by users 
from the source programs.
Refer to the GR@PPA 2.8 manual~\cite{Odaka:2011hc} 
for the installation of these external libraries.

We can choose HBOOK or RBOOK as the histogramming tool 
in the built-in sample programs.
We need to install CERNLIB libraries for using HBOOK, 
and ROOT libraries for using RBOOK.
Although the binary distributions of ROOT are available from its official 
Web site and/or standard repositories of the Linux distributions,
the availability of CERNLIB is limited on some platforms 
as the official support by CERN has already been terminated.
We could not find the binary distributions of CERNLIB for CentOS 7 and 
Fedora 24, while it could be installed from the EPEL repository on SLC 6 
and from the standard repository on Ubuntu 16.04.
In the latter two cases, the installed libraries need to be linked 
with the "static" option.
See the comments in the sample programs for more details.
We provide only the HBOOK version for the external sample programs.
However, if necessary, the conversion to RBOOK would be straightforward 
since HBOOK is used only for histogramming.

\begin{table}[t]
\caption{CPU time consumption of the external "matched diphoton + 2 jets" 
sample program.
The time that is consumed for the cross-section integration by BASES 
and that for the event generation by SPRING are separately presented 
for each subprocess.
The achieved integration accuracy is also presented.
The results were obtained using 2.8-GHz CPU computers
with the SLC 5 operating system.}
\label{tab:perf}
\begin{center}
\begin{tabular}{l c c c c c c c}
\hline
Subprocess & $aa2j$ & $aa1j$ & $aa0j$ &  $a2j$ &  $a1j$ & $qcd2j$ & $gg2aa$\\
\hline
\\
BASES & 95 h 25 m & 1 h 3 m  & 3 m 13 s & 16 h 17 m & 50 m 18 s 
& 34 h 41 m & 12 m 0 s  \\
accuracy (\%) & 0.79 & 0.17 & 0.14 & 0.37 & 0.28 & 0.19 & 0.98  \\
\\
SPRING (s/event) & 9.1 & 0.015 & 0.0010 & 0.49 & 0.024 & 0.51 & 0.19 \\
\\
\hline
\end{tabular}
\end{center}
\end{table}

The CPU time consumed for the execution of the external 
"matched diphoton + 2 jets" sample program is summarized 
in Table~\ref{tab:perf}.
The test was carried out on Intel Xeon 5560 (2.8 GHz) machines
operated with SLC 5.
The program size was approximately 4.5 MB.
The virtual memory size was 20 MB, 
and the physical (resident) size was approximately 4 MB 
including the shared memory of about 2 MB during the execution.
The CPU time consumption for the integration/optimization by BASES 
and that for the event generation by SPRING are separately presented 
in the table for each subprocess: $\gamma\gamma$ + 2 jets ($aa2j$), 
$\gamma\gamma$ + 1 jet ($aa1j$), $\gamma\gamma$ + 0 jet ($aa0j$), 
$\gamma$ + 2 jets ($a2j$), $\gamma$ + 1 jet ($a1j$), QCD 2-jets ($qcd2j$), 
and the loop-mediated $gg \rightarrow \gamma\gamma$ ($gg2aa$).
The results are strongly dependent on the subprocess.
In general, 
the time consumption increases as the particle multiplicity increases.
Additional time is consumed for the single-$\gamma$ production processes 
($a2j$ and $a1j$) and $qcd2j$, 
because the radiation of energetic photon(s) is required in PS.

The results for $gg2aa$ is markedly worse 
although it is a simplest two-body production process.
This is because the PS energy scale is set to be very large 
in this subprocess in order to include hard radiation effects.
The constraint on the initial hard interaction had to be significantly 
relaxed in order to avoid any bias to the final sample, 
as the kinematical alteration by PS is very large.

We carried out the test by using the built-in PDF with control parameters 
that had been optimized in a previous study with another PDF.
Better integration accuracy can be achieved 
by increasing the {\tt NCALL} parameter, 
although the integration time will increase in proportion to {\tt NCALL}.

\section{Summary}
\label{sec:summary}

Our understanding of photon production processes in hadron collisions is 
still insufficient. 
We have explicitly shown large contributions from parton-associated processes 
to the direct (single) photon and diphoton (double photon) production 
in proton-proton collisions 
in our previous studies~\cite{Odaka:2012ry,Odaka:2015uqa,Odaka:2016tef}.
We release the simulation programs used in these studies 
as a program package, GR@PPA 2.9.
The included programs consistently combine ME-based simulations for 
parton-associated processes up to two partons with PS simulations 
that reproduce the QCD multiple radiation and QED fragmentation.

The simulation method and resultant physics performance 
have already been described in the previous articles.
We provided instructions for the installation and execution 
of the programs in this article.
Since the previous release of GR@PPA, GR@PPA 2.8, 
some parameters have been newly introduced and some have been extended 
in order to control the new functions, namely the LabCut framework, 
forced photon radiation, and soft-gluon correction.
In addition, programs for defining the energy scales have been modified 
to accomplish exact matching between the processes to be combined.
The LabCut framework provides a new scheme for event selection based on 
the event information after the application of PS.
The selection condition is reflected by the cross section integration 
and event-generation optimization.

We have confirmed that the programs can be built and executed 
on various Linux platforms.
The practical performance of the program was presented for 
a sample program of the "matched diphoton + 2 jets" process.
The CPU time consumption is strongly dependent on the composing subprocess.
In general, longer CPU time is required for larger multiplicity processes.
The requirements of additional photon radiation in the PS simulation 
result in additional consumption of CPU time.

\section*{Acknowledgments}

This work has been carried out as an activity of the NLO Working Group, 
a collaboration between the Japanese ATLAS group and the numerical analysis 
group (Minami-Tateya group) at KEK.
The authors wish to acknowledge useful discussions with the members, 
especially S. Tsuno and J. Fujimoto of KEK, and K. Kato of Kogakuin U.


%% References


%\bibliographystyle{elsarticle-num}
%\bibliography{grappa29}






% \end{thebibliography}

\end{document}